# Application of Machine Learning in Identification of Best Teaching Method for Children with Autism Spectrum Disorder

Zarin Tassnim Zoana
*Dept. of Computer Science and Engineering*
Brac University
Dhaka, Bangladesh
zarin.tassnim.zoana@g.bracu.ac.bd

Mahmudul Wahed Shafeen
*Dept. of Computer Science and Engineering*
Brac University
Dhaka, Bangladesh
mahmudul.wahed.shafeen@g.bracu.ac.bd

Nasrin Akter
*Dept. of Computer Science and Engineering*
Brac University
Dhaka, Bangladesh
nasrin.akter2@g.bracu.ac.bd

Tanvir Rahman
*Dept. of Computer and Information Sciences*
College of Engineering,
University of Delaware
Delaware, United States of America
rtanvir@udel.edu

*Abstract*—A good teaching method is incomprehensible for an autistic child. The autism spectrum disorder is a very diverse phenomenon. It is said that no two autistic children are the same. So, something that works for one child may not be fit for another. The same case is true for their education. Different children need to be approached with different teaching methods. But it is quite hard to identify the appropriate teaching method. As the term itself explains, the autism spectrum disorder is like a spectrum. There are multiple factors to determine the type of autism of a child. A child might even be diagnosed with autism at the age of 9. Such a varied group of children of different ages, but specialized educational institutions still tend to them more or less the same way. This is where machine learning techniques can be applied to find a better way to identify a suitable teaching method for each of them. By analyzing their physical, verbal and behavioral performance, the proper teaching method can be suggested much more precisely compared to a diagnosis result. As a result, more children with autistic spectrum disorder can get better education that suits their needs the best.

## I. Introduction

Autism Spectrum Disorder or ASD is a disorder of neurological development that mostly occurs in the early stage of a child's life, that can be distinguished by a persistent deficiency in social interaction and communication and repetitive behaviors[1]. According to the World Health Organization, about one in 270 people globally is estimated to have ASD[2]. And 31 percent of children with ASD have an intellectual disability, that is, their intelligence quotient (IQ) is less than 70[3]. That is why students with ASD do not reach the same academic outcomes as opposed to other students[4]. Article 26 of the Universal Declaration of Human Rights states that everyone has the right to education[5]. That means all types of students, like children with disabilities and fromminority groups from all corners of society have the right to get educated, and that includes children with autism spectrum disorder. In the last two decades, there has been a huge increase in researches on autism spectrum disorder[6] and following this, the tendency of including autistic children in the general and specialized education system has also ascended.

However, education for children with autism spectrum disorderis far from perfect. Students with ASD require a special form of education and are thus considered as students with "special educational needs" or SEN[7]. Arranging this SEN is mostly hard not only due to the student lacking communication and interaction skills, but also due to lack of collaboration between teachers and parents, and a very limited amounts of practical suggestions and proper resources to guide them[7].

Each year more researches are done on autism spectrum disorder and specialists are able to diagnose children with ASD much more successfully than before. In the last decade, the number of children diagnosed has increased an incredible five times[6]. Based on their diagnosis, the children are categorized into their own sections of special education. But quite often, they are rejected in general schools, and even when they get their primary education in a specialized school, later in life, they drop out during their higher studies, due to anxiety and poor academic achievement.

Most researches on autism spectrum disorder are done using large-scale statistics. Since diagnosis is done based on statistics and not persistent data, there might be some inaccuracies. And thus, identifying their special educational needs is not so accurate as well. But this can be done more efficiently using Machine Learning algorithms that can solve large amounts of variables more precisely.

## II. RELATED WORK

This section will evaluate previous relevant machine learning work in the context of determining the optimum teaching method for autistic children. We examine the different strategies utilized to accomplish the major conclusions and explain how matching learning has its own set of challenges and limitations this regard. Since the past decade machine learning approaches have been very convenient to diagnose ASD in children. For instance, Virginia Tech Center along with Virginia Tech Institute for Society, Culture and Environment (ISCE), did research where they applied fast Artificial Neural Network (fANN) technique using data from 14,995 infants (16-30 months, 46.51% male) taking twenty inputs and produce an assertive or negative output if the kid has ASD. The sample was then clustered into groups according to their race, sex, and maternal education (i.e., mother has completed Associate Degree or not). The results yielded 99.72% accurate, with 99.92% accuracy for white children, and 99.79% for those black. The results were 99.64% correct for the boys, while it was 99.95% for the girls. In case of maternal education, the results were 99.75% correct for mothers having completed the degree and 99.70% for those who have not[8]. However, very recently, focus has been shifted to use of technology in educational needs of special children. Department of Computer Science and Institute of Education, University College London used machine learning techniques to figure out how students respond to various forms of communication used by their teachers inside controlled classroom conditions considering their specific attributes[9]. There is a lot of space for further researches in this field of taking machine learning approaches to find the proper teaching method for children with ASD.

## III. BACKGROUND

Supervised Learning is one of the three types of machine learning, which in turn is a subcategory of artificial intelligence. This type of machine learning is known for using labeled datasets to determine patterns, create classification of the data, and calculate estimated results as precisely as possible. It takes a set of data to learn patterns and educate particular models and then yields the desired output. The model later incorporates more inputs to the datasets and compare outputs, which allows it to learn over time and become more accurate in giving outputs. The model measures its accuracy through the loss function, updating itself until the error has been adequately reduced to the point where it is negligible[10]. We decided to use Supervised learning algorithm as it permits collecting information and produces output from past encounters. It makes a difference to optimize execution criteria with the assistance of past experience.

Naive Bayes is a probabilistic classifier, that is based on the Bayes hypothesis with presumption of independence among predictors. Bayes hypothesis depicts the probability of an event to occur by comparing it to the previous records of the event's probability. It is a classification approach that uses the theory of class conditional independence from the Bayes Hypothesis. This implies that the existence of one feature does not impactthe existence of another within the probability of a given result, and each predictor has a rise to impact on that result[11]. Confusion Matrix represents the performance estimation of two or more classes of outputs from classification done with a machine learning model. It is a matrix used to portray the execution of a classification model by comparing a set of test data, whose true values are known, and a similar set of predicted data. Confusion matrix can be very useful for measuring Review, Accuracy, Specificity, Precision, and most imperatively AUC-ROC bends[12].

Random Forest is a popular supervised machine learning algorithm. It can be used as a Classification model as well as a Regression model. It is based on the notion of supervised learning, which combines multiple classifiers and solves complex problems and moves forward to execute the model. It is a classifier that combines a bunch of decision trees on different parts of a particular datasets and uses the normal to improve predictive precision of the dataset. Rather than relying on a single decision tree, the random forest takes the output from each decision tree and predicts the final output based on the majority of predictions. The larger number of trees within the forest, the more precise it is and avoids any complication of over fitting[13].

Decision tree is known as an effective algorithm in the case of prediction as well as classification. It can be used in the supervised learning method. A Decision tree looks like a flowchart type of tree structure and on each inner hub indicates a test on an attribute. In the decision tree algorithm, each branch represents an output of the given input-based testing, and each leaf node, also we call it terminal node holds a class label. In a Decision tree, there are two nodes that are being used, first one is Decision Node and the other one is Leaf Node. Mainly, Decision nodes are utilized to make any decision. Decision tree algorithm has numerous branches, while its Leaf nodes are the output of those decisions and do not contain any further branches[14]. On the premise of features of the given datasets, here almost all decisions are performed. It is a graphical representation for getting all the conceivable outputs to a problem based on given conditions. There is a specific reason to call it a decision tree because it begins with the root node that extends on further branches comparable to a tree and builds a tree-like structure. A decision tree basically inquires, and according to the reply (Yes/No), it assists parting the tree into sub trees.

The K-Nearest Neighbor technique assumes that the unused data and accessible cases are close in proximity, and places the new data in the category that is most similar to the available categories. The K-NN algorithm saves all available data and categorizes unused data points depending on their proximity. This means that as new information appears, it may be quickly sorted into a suitable category using the K-NN algorithm. The K-NN algorithm can be used for both regression and classification, but it is more commonly used for classification tasks. It is also known as a lazy learner algorithm since it does not learn from the training set right away; instead, it saves the datasets and performs an action onit when it comes time to classify it. At the training stage, the K-NN algorithm simply saves the datasets, and when it receives new data, it classifies it into a category that is significantly more comparable to the new data[15].

IV. PROPOSED APPROACH

The goal of the proposed model is to apply machine learning to determine the best teaching method for children with ASD. To accomplish said goal, the model needs to design a process that accepts data as an input, process it, apply the machine learning algorithms and find the best fit. The figure below provides an effective view of the model design. First, we have to collect data relevant to the research. The input data preprocessing stage is concerned with dropping unnecessary data and convert the datatype of necessary data to make it easy to process in the model. The preprocessed input data goes into the train and test splitting stage and build two segments one used to train the model and the other used to compare with the predicted outputs. After that, selected machine learning algorithms are used and run on the preprocessed input data for selecting the best fit algorithm and apply them in finding the appropriate teaching method for children with ASD.

Among the many education methods specialized for children with autism, six teaching methods have been selected for the project. The first one is Technology-aided instruction and intervention, which is a learning system that uses technology as its central feature. This is designed for children who need some extra time to go over an exercise and adjust to their own pace of learning[16]. As a result, teachers and other students needing less support can continue ahead. Secondly, Antecedent based intervention is another evidence-based practice that identifies what causes interference to a child and modifies the environment to remove said interfering behavior of the child so that he/she can regain focus on the exercise[17].

Thethird teaching method used in the project is Pivotal response training. This is a therapy type training system that increases a child's motivation to learn, start communication with someone, and monitor their own behaviors[18].

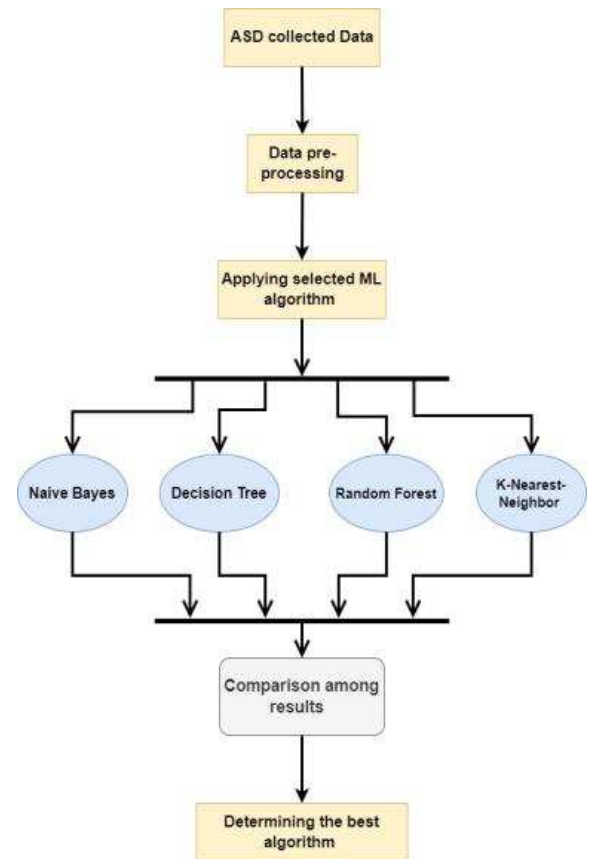

Fig. 1.  *Proposed Approach*

On the fourth spot, we have Peer-mediated instruction and intervention. This method involves another child without disabilities to take on a role in the teaching beside the teacher or therapist[19]. The peer not only plays the role of a tutor, but also teach critical social skills along the way. Next is the Picture Exchange Communication method, which encourages verbally challenged children to use visual symbols to communicate with parents, teachers and peers. It is a kind of complementary and alternate system that is used for intentional and functional communication to tell what they want or need[20]. Lastly, we have selected Task Analysis, which breaks down complex tasks into sequential smaller steps. This teaching method is applicable for individuals who find even simple tasks to be difficult and give up[21].

*A. Input Data*

As for input data to determine the best teaching method for children with autism, it was quite difficult to obtain data, since such data are confidential, and very few researches have been done on this. We are using supervised Machine Learning method for our research and so, we have searched for labeled data. Since there was no dataset large enough to be used for Machine Learning, we have used a total 4 datasets for our research and run our algorithms on a merged dataset with same attributes. We have used the dataset named 'Autistic Spectrum Disorder Screening Data for Toddlers' documented by Dr. Fadi Fayez Thabtah, Principal Lecturer of DigitalTechnology department of Manukau Institute of Technology in New Zealand[22].Download Link:
https://www.kaggle.com/fabdelja/autism-screening-for-toddlers?select=Toddler+data+description.docx
Second dataset that we used named Behavior Analysis of Autism.Download Link:
https://www.kaggle.com/iashiqul/behavior-analysis-of-autism

The other two datasets we have used the data set named Autism screening child consists of two version. It has 2249 instances, 18 attributes.Download Link:
https://www.kaggle.com/basmarg/autism-screening-child-two-version

Datasets regarding clinical or screening autism spectrum disorder are presently very limited and also generic. That is why we used this merged dataset because it is based on influential features that can be useful for further research in improving the classification process and determining ASD cases.The dataset contains categorical, binary and continuous data. The information is focused on medical, health and social science areas. The dataset holds 3043 samples of different children with autism spectrum disorder. The dataset contains 12 integer type data including case number, closed questions whose answers are represented by binary 1's and 0's and Q-chat-10 score. The other 7 data are object type. The total size of our merged dataset is 451.8 kB.

We have used a merged dataset that consists of 4 individual datasets that has 3043 instances in total. We decided to use this dataset because the problem with scarcity of data is a crucial drawback, as a good amount of data is integral to any project done with machine learning algorithm. If a dataset is inadequate, it might as well be the agent of poor performances in the project. Very often, drawbacks like these are the main reason why significant machine learning projects remain unaccomplished. Most supervised learning algorithms are strongly reliant on the amount of training data provided. It can be challenging to create large enough training datasets in many circumstances. With smaller datasets algorithms tend to learn the detail of the noises in the training data to such extent, that the model's performance on newer data can have a negative impact. In general, the smaller the dataset, the better it is to use the simpler the machine learning algorithm. Small data requires low-complexity models in machine learning in order to prevent overfitting the model to the data. The Naive Bayes for example, is one of the most basic algorithms, and as a result, it's performance on learning from comparatively minor dataset can indeed be remarkably well. Furthermore, other simpler algorithms such as decision tree, random forest and K-NN can also learn really well from small datasets. These algorithms are essentially better than more complicated algorithms in case of learning from smaller datasets, as they actually try less to learn from exceptional cases or noises.

*B. Data Preprocessing*

Now that we have the merged dataset, we initially tried to run four Machine Learning algorithms through the dataset without any kind of pre-processing techniques. That means, we only got to use the integer datatypes for training the algorithms. These were the A1-A10 answers and Qchat-10 score. However, three out of four algorithms gave 100 percent accuracy and the other one gave 99.8 percent accuracy. This case can be interpreted as such that there was no point in using Machine Learning and this could be done only with a bunch of if-else conditions. That is when we decided to include more features of object datatype. So, at first, we used Label Encoding pre-processing method to convert the object datatypes into integer datatype so that the algorithms can read those features as well. By doing so, we trained the algorithms with 16 features instead of 11 features done previously. With more input data, a machine learning algorithm can learn patterns more efficiently.

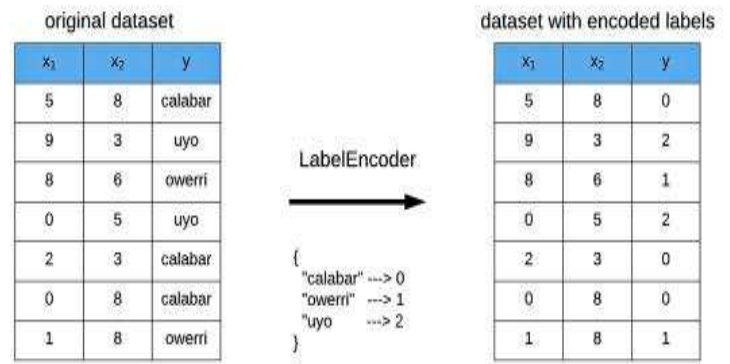

Fig. 2. *Label Encoding*

However, due to Label Encoding, some data became unbalanced, so after splitting the training and testing sets, we used Feature Scaling pre-processing on the training set to achieve a higher accuracy rate.

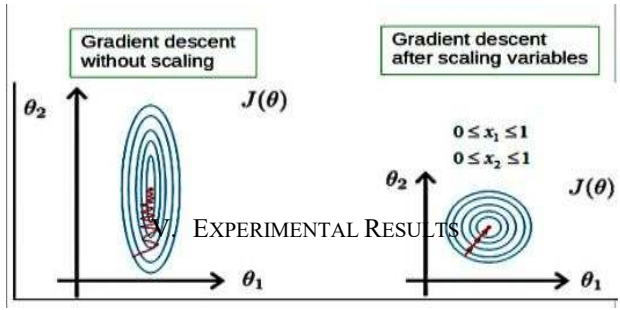

Fig. 3. *Label Encoding*

The objective of our project is to use a Machine Learning Algorithm that will learn from patterns in autistic traits, sex, and other factors responsible for autism and then predict a preferred educational method for an individual. Due to its access to many built-in libraries such as numpy, pandas, matplotlib, sklearn, the most preferred language Python has been used in the project. The project is done in Google Colaboratory. The datasets used for the project have been stored in Google Drive in the same folder as the Colab Notebooks.

*A. Inplementation*

The project file starts by importing the necessary libraries and frameworks. At the next step Google Drive is mounted with Colab and the dataset is read by the project file using pandas library, which is the best way to handle large data frames in a Python program. For simplicity and efficiency, we preprocess the dataset with the help of two methods. We used Label Encoding to include more features to train the algorithms. The binary data A1 to A10 and Q-chat-10 Score are kept as is, while the rest of the columns that had object datatype are converted to integer datatype and added as new columns. We also used Feature Scaling on the training data to balance out irregularly encoded data.

Another new column namely "Preferred Education" is added using the numpy library. This column consists of integers 1 to 6 to represent the six special teaching methods, and 0 to represent that no special education is required. These six teaching methods are selected based on some conditions set upon the binary datatype autism traits A1-A10, as followed:

TABLE I. PREFERRED EDUCATION BASED ON A1 TO A10 SCORES

| Technology-aided Instruction | Antecedent based Intervention | Pivotal Response Training | Peer-mediated Instruction and Intervention | Picture Exchange Communication | Task Analysis |
|---|---|---|---|---|---|
| A5 = 1 | A6 = 1 | A1 = 1 | A5 = 1 | A2 = 1 | A7 = 1 |
| A9 = 1 |  | A8 = 1 | A4 = 1 | A9 = 1 |  |
| A10 = 0 |  |  | A3 = 1 |  |  |

*Yes = 0, No = 1*

*B. Result*

At first, the columns of dataset are divided to 2 sets, one as independent variable x, which includes A1, A2, A3, A4, A5, A6, A7, A8, A9, A10, Q-chat-10 Score, Age Mons enc Sex enc, Ethnicity enc Jaundice enc, and Family mem with ASD enc and Class/ASD Traits enc. The other set of dependent variable y consists of just the 'Preferred Education' column. In the next step we split both x and y part of the dataset to the training and testing set. To be more accurate, the testing part is set to be 0.05, i.e. 5% of the total data. That means that the algorithm will learn from 95% of the data. It will find patterns from the x and y of the training set and then predict y from the x of the testing set. Finally, the predicted y values are checked with y values of the testing set to calculate accuracy of the algorithms. Here are the Confusion matrices and Accuracy score of all four algorithms.

TABLE II. ACCURACY SCORE

| Algorithms | Accuracy |
|---|---|
| Naïve Bayes | 90.85% |
| Decision Tree | 98.69% |
| Random Forest | 98.69% |
| K – Nearest Neighbor | 94.12% |

TABLE III. PRECISION SCORE

| Algorithms | Accuracy |
|---|---|
| Naïve Bayes | 91.70% |
| Decision Tree | 97.55% |
| Random Forest | 99.10% |
| K – Nearest Neighbor | 92.64% |

TABLE IV. RECALL SCORE

| Algorithms | Accuracy |
|---|---|
| Naïve Bayes | 95.24% |
| Decision Tree | 98.80% |
| Random Forest | 97.95% |
| K – Nearest Neighbor | 91.88% |

TABLE V. F1 SCORE

| Algorithms | Accuracy |
|---|---|
| Naïve Bayes | 92.19% |
| Decision Tree | 98.08% |
| Random Forest | 98.48% |
| K – Nearest Neighbor | 92.10% |

It is easily understandable that both Random Forest and Decision Tree have the highest accuracy of approximately 98.69 percent. But Random Forest has the highest precision score of 99.09 percent and Decision Tree has the highest recall score. Now the highest F1 score of Random Forest, i.e. 98.47 percent breaks the tie. Thus by comparing the accuracy, precision, recall and F1 score, the best fit for our approach has to be Random Forest.

VI. CONCLUSION

It is effectively necessary to find a suitable teaching method for the children having autism spectrum disorder because every child with ASD has different characteristics and they should have specific education which is appropriate to their needs. Based on a report of World Health Organization (WHO), among every 160 children, at least one has been diagnosed with autism spectrum disorder (ASD). According to Paulette Delgado in her article Autism Spectrum Disorder (ASD) in Education, Schools fail to fulfill the expectations of guardians by not recognizing or supporting the requirements of their children who need special treatment[23]. Social interactions are everywhere and changing continuously in schools. Moreover, some activity in the classroom may be suitable for generalstudents but it can be unfitting for the children with ASD. One the other hand some social prompts or signs which indicate a child to change their certain behaviors are usually difficult to follow for an autistic child. There is confusion between teachers on how to treat an autistic child with special educational needs as there is a deficiency of the depth of knowledge. This might have a negative consequence on their education.

Earlier studies have anticipated the challenges of teachers regarding the right education method. Not much research has been done previously to implement machine learning in identification of proper education needs of a child with ASD[24]. So, this research can be helpful for teachers, parents as well as education institutions to determine an effective learning method for an autistic child. We have suggested six education systems according to their necessity. After that four machine learning algorithms named Naïve Bayes, Random Forest, Decision Tree and K-Nearest Neighbors have been used in this research. According to the result we have found that the Random Forest algorithm gave approximately 98.69% accuracy, 99.10% precision, 97.95% recall and 98.48% f1 score, which is the best fit in finding the most appropriate teaching approach based on the children's characteristics.

ACKNOWLEDGMENT

The authors would like to thank Tanvir Rahman, Lecturer of CSE Department, BRAC University for his unwavering support, guidance, and encouragement throughout the process. We would also like to express our gratitude to Brac University for providing us with the chance and support we needed to complete this research.